\documentclass[letterpaper,twocolumn,reprint,amsmath,amssymb,aps,prl]{revtex4}
\usepackage{lmodern}

\usepackage[T1]{fontenc}
\usepackage[latin9]{inputenc}
\usepackage{color}
\usepackage{amsmath}
\usepackage{graphicx}
\usepackage[unicode=true,pdfusetitle,
 bookmarks=true,bookmarksnumbered=false,bookmarksopen=false,
 breaklinks=false,pdfborder={0 0 0},pdfborderstyle={},backref=false,colorlinks=true]
 {hyperref}

\makeatletter

\@ifundefined{textcolor}{}
{%
 \definecolor{BLACK}{gray}{0}
 \definecolor{WHITE}{gray}{1}
 \definecolor{RED}{rgb}{1,0,0}
 \definecolor{GREEN}{rgb}{0,1,0}
 \definecolor{BLUE}{rgb}{0,0,1}
 \definecolor{CYAN}{cmyk}{1,0,0,0}
 \definecolor{MAGENTA}{cmyk}{0,1,0,0}
 \definecolor{YELLOW}{cmyk}{0,0,1,0}
}

\makeatother

\begin{document}
\title{Prototype of a Phonon Laser with Trapped Ions}
\author{Chen-Yu Lee, Kuan-Ting Lin, and Guin-Dar Lin}
\affiliation{Department of Physics and Center for Theoretical Physics, National
Taiwan University, Taipei 10617, Taiwan\linebreak{}
Center for Quantum Science and Engineering, National Taiwan University,
Taipei 10617, Taiwan}
\begin{abstract}
We propose a tunable phonon laser prototype with a large trapped ion
array, where some of the ions are effectively pinned by optical tweezers,
thus isolating a subset of ions that mimics an acoustic cavity used
as a phonon lasing resonator. The cavity loss can then be controlled
by the tweezer strength and the ``wall thickness'', the number of
pinned ions for isolation. We pump the resonator by applying blue-sideband
lasers, and investigate the lasing dynamics of the cavity modes such
as threshold behavior, population distribution, the second-order coherence,
and line-narrowed spectrum. This scheme can be generalized to resonators
consisting of multiple cavity modes formed by a few ions, where we
demonstrate mode competition and synchronization as lasing modes have
developed.
\end{abstract}
\maketitle
\noindent \textit{Introduction.} \textendash{} Laser technology has
been one of the most important ingredients in contemporary scientific
research, industry, and consumer electronics. Many remarkable properties
of an optical laser like quantum coherence and capability to travel
over a long distance make it a unique tool in applications of modern
quantum engineering and communication. Recently, acoustic analog of
lasing phenomena has drawn growing interest. This line of research
extends our understanding from ordinary quantum optics to other physical
degrees of freedom for their mathematical frameworks share essential
similarities. Acoustic waves are much slower than the light and hence
of shorter wavelength, thus providing an opportunities for precise
phase control. Further, the interaction processes between an atom
and phonons can be deterministic without additional waveguides, which
is very useful for quantum computing \citep{Cirac1995}.

The first phonon laser was realized in a trapped ion system driven
by optical forces \citep{Vahala2009}, where a single ion presents
self-sustained oscillation beyond a threshold gaining energy from
optical sources. Since then, many proposals have been studied and
demonstrated in similar platforms \citep{Knunz2010,Ip2018} and others
such as quantum dots \citep{Kabuss2012,Kabuss2013,Khaetskii2013},
and optomechanical systems \citep{Grudinin2010,Beardsley2010,Khaetskii2013,Mahboob2013,Kemiktarak2014,Jing2014,Zhang2018,Jiang2018,Pettit2019,Sheng2020}.
Many intriguing properties of ordinary lasers have also been reported
in phonon systems, including oscillation threshold \citep{Vahala2009,Grudinin2010,Khurgin2012,Mahboob2013,Kemiktarak2014,Zhang2018,Pettit2019,Sheng2020},
Poissonian distribution \citep{Pettit2019}, linewidth narrowing \citep{Grudinin2010,Beardsley2010,Khurgin2012,Mahboob2013,Zhang2018,Pettit2019},
injection locking \citep{Knunz2010,Ip2018}, and mode competition
\citep{Kemiktarak2014,Sheng2020}. Most of the schemes, however, are
based on sophisticated designs of architecture and cannot be easily
scaled up to include more modes. The parameters of the lasing resonator
are typically fixed upon fabrication, limiting the opportunities of
exploring rich phonon physics.

\begin{figure}[t]
\begin{centering}
\includegraphics[width=8.5cm]{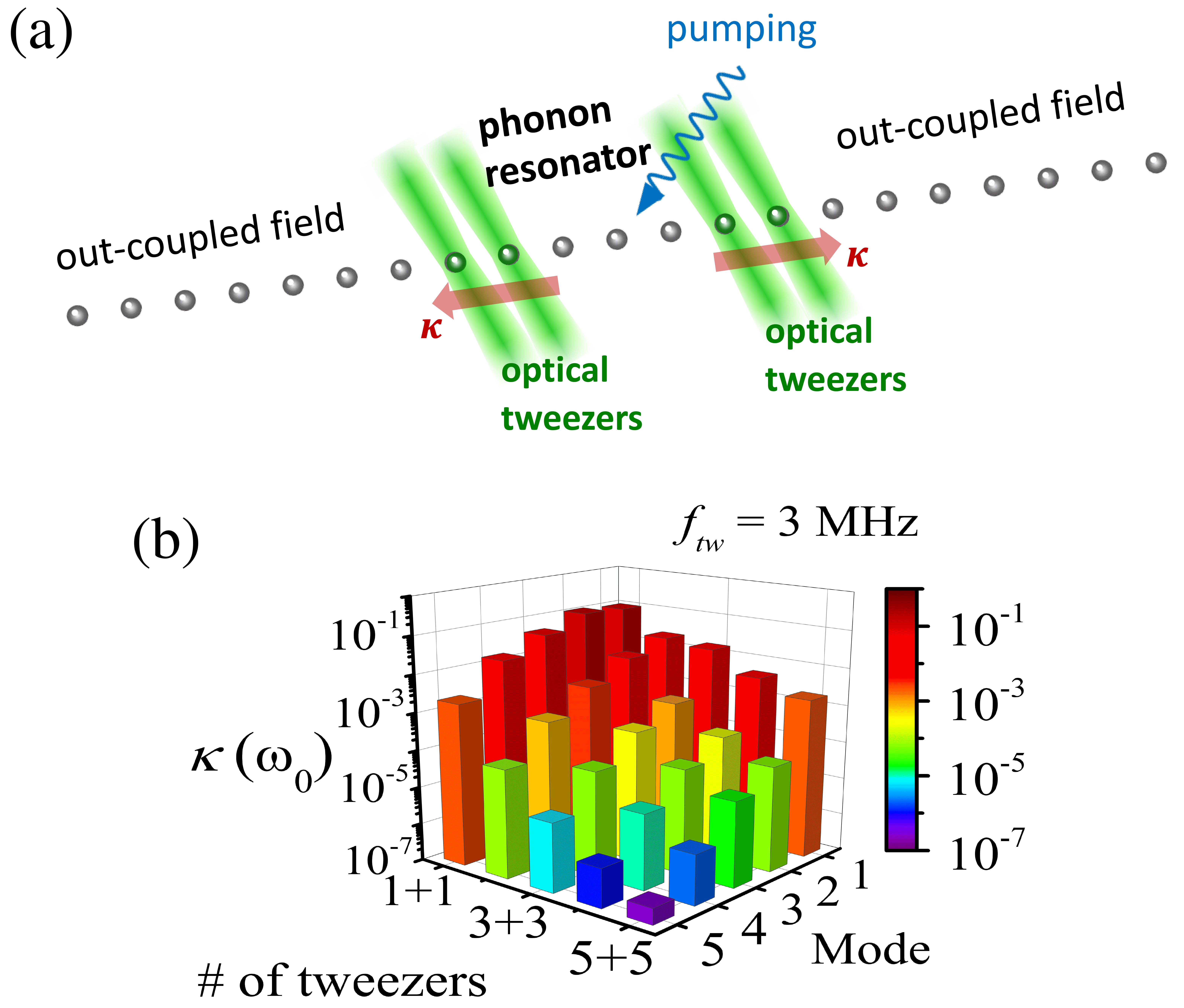}
\par\end{centering}
\caption{(a) Architecture of an effective phonon resonator constructed on a
large-scale ion crystal. The tweezered ions act as partial mirrors
of the resonator. (b) Mode decay rates against varied wall thickness
of an $N_{S}=5$ cavity. The collective normal modes are ordered according
to their frequencies. (1: lowest; 5: highest) These results are based
on calculation considering an $N>2000$ $^{40}$Ca$^{+}$ ion crystal
with ion separation $7$ $\mu$m.\label{fig:architecture}}
\end{figure}

Here, we propose a prototype of a tunable phonon laser based on a
large uniform ion crystal and optical tweezers. Such an ion crystal
can be constructed in a long Paul trap \citep{Shen2020}, microtraps
or Penning microtrap arrays \citep{Cirac2000,Ratcliffe2018,Jain2020}.
We apply optical tweezers on one ion or a few in a row so that they
form a ``wall'' for acoustic waves. We then consider a small subset
of ions being contained by two such walls, forming an effective phonon
cavity as shown in Fig.~\ref{fig:architecture}(a). Note that, for
a uniform ion crystal, the frequency scale characterizing momentum
exchange between adjacent ions is $\omega_{0}\equiv[e^{2}/(4\pi\epsilon_{0}md_{0}^{3})]^{-1/2}$,
obtained by matching the energy scales of local oscillation and mutual
Coulomb interaction, where $e$ is the charge carried by an ion of
mass $m$. For ion separation $d_{0}$ about a few microns, $\omega_{0}$
is of order of magnitude about hundreds of kilohertz to a few megahertz.
Therefore, the application of optical tweezers of strength larger
than $\omega_{0}$ by roughly an order can significantly modify the
motional spectrum, resulting in a collection of local modes formed
within the ``cavity''. This remarkable feature allows us to view
the system as a phonon laser resonator.

The beauty of this proposed scheme is its simplicity and flexibility
of reconfiguration. Note that optical tweezers can be switched on
and off easily at a timescale of nanoseconds, without altering the
spatial equilibrium of the array. Further, the effective cavity is
scalable in size on demand. The reflectivity of a partial mirror can
be tuned by varying tweezer frequencies and/or numbers of tweezers.
In following discussion, we demonstrate the idea by looking into the
lasing dynamics assuming the ion array is only Doppler cooled.

\textit{Model.} \textendash{} We consider $N_{S}$ ions within the
effective resonator, and only focus on $N_{S}$ longitudinal modes.
This is because the rest of the array contribute to a broader dispersion
band of bath in the longitudinal modes than the transverse ones. A
wall thickness $w$ is represented by the number of tweezered ions.
We use $w_{L}+w_{R}$ to denote the thicknesses of the left and right
walls. Without altering the general conclusion, in this work we look
at the symmetric cases with $w_{L}=w_{R}$. The number of bath ions
is $N_{B}=N-w_{L}-w_{R}-N_{S}$, where $N$ is the total number of
ions. It is assumed that $N_{B}\gg N_{S}$ such that the discrete
bath spectrum approximate a continuous band and is treated Markovian.

The motional Hamiltonian is described by $H_{m}=\sum_{i}p_{i}^{2}/(2m)+\sum_{i,j}A_{ij}z_{i}z_{j}$,
where $z$ labels the longitudinal direction, $z_{i}$ is coordinate
operator with respect to the equilibrium position of the $i$th ion,
and the associated momentum $p_{i}$. The coupling matrix elements
$A_{ii}=\nu_{i}^{2}+\nu_{i}^{{\rm ot}2}+\sum_{l=1,l\neq i}^{N}2/|u_{i}-u_{l}|^{3}$
and $A_{ij}=-2/|u_{i}-u_{j}|^{3}$ for $i\neq j$, where $u_{i}$
is the equilibrium $z$ position of the $i$th ion in units of $d_{0}$
\citep{Zhu2006}. Also, $\nu_{i}=\omega_{i}/\omega_{0}$ and $\nu_{i}^{{\rm ot}}=\omega_{i}^{{\rm ot}}/\omega_{0}$
are dimensionless frequencies introduced by the global trap and optical
tweezers, respectively. Here, we use the trap configuration discussed
in \citep{Shen2020}, a large linear Paul trap with a box-like potential
so that $\omega_{i}\approx0$ except those near the edges, where the
exact form of the potential profile needs to be computed with care.
We also assume that optical tweezers are applied transversely to the
array so $\omega_{i}^{{\rm ot}}>0$ for tweezered ions; otherwise,
$\omega_{i}^{ot}=0$.

By dividing the whole array into the system $C$ {[}Appendix~\ref{sec:a.kappa}{]}
and the bath $B$, and using the phononic field operator representation,
the motional Hamiltonian can be recast into 
\begin{align}
H_{m} & =\sum_{q\in C}\hbar\omega_{q}a_{q}^{\dagger}a_{q}+\sum_{k\in B}\hbar\omega_{k}a_{k}^{\dagger}a_{k}\nonumber \\
 & +\sum_{q\in C,k\in B}g_{qk}\left(a_{q}a_{k}^{\dagger}+\text{H.c.}\right),\label{eq:Hm}
\end{align}
where $a_{q}$ ($a_{k}$) and $a_{q}^{\dagger}$ ($a_{k}^{\dagger}$)
are phononic annihilation and creation operators, respectively, of
the $q$th ($k$th) normal modes of $C$ ($B$). H.c. stands for the
Hermitian conjugate. The system-bath mode coupling $g_{qk}=g_{kq}^{T}=\frac{\hbar}{2m}\sum_{i\in C,j\in B}U_{C,qi}^{T}A_{ij}U_{B,jk}/\sqrt{m^{2}\omega_{q}\omega_{k}}$,
where the matrices $U_{C}$ and $U_{B}$ diagonalize the corresponding
submatrices in matrix $\mathbf{A}\equiv[A_{ij}]$. Since $N_{B}\gg N_{S},$the
excitation within the cavity can dissipate to the bath's degrees of
freedom, and only return after a long time $\sim N_{B}\omega_{0}^{-1}$.
This timescale is given by the elapsed time of motion propagation
to the edge and back. Before the revival happens, the dissipation
of the cavity modes can be characterized by the decay rates
\begin{align}
\kappa_{q} & \approx2\pi\bar{g}_{qq}^{2}\rho_{B}(\omega_{q})\label{eq:kappa}
\end{align}
according to the standard Fermi golden rule approach. Here, we have
taken the continuum limit for $B$ and numerically computed the density
of states $\rho_{B}(\omega)$ of the bath. We obtain $\bar{g}_{qq}^{2}$
by coarse-graining $|g_{qk}|^{2}$ over a small range of $\omega_{k}\approx\omega_{q}$,
that is, $\bar{g}_{qq}^{2}\equiv\langle|g_{qk}|^{2}\rangle_{\omega_{k}\approx\omega_{q}}.$
Note that this approach is valid as long as the Markovian bath assumption
holds. For a large but finite $N\sim\mathcal{O}(10^{3})$, we also
numerically check the time evolution of the cavity mode population,
and extract the decay rates by fitting to an exponential profile.
Our results show very good agreement with Eq.~(\ref{eq:kappa}) {[}Appendix~\ref{sec:a.kappa}{]}.

It can be expected that increasing the wall thickness and/or tuning
up the tweezer strength help isolation of the cavity from the rest
of the ion crystal, and therefore the decay rate of a cavity mode
decreases. This provides an extremely convenient way to setup the
cavity because the state-of-the-art strongest tweezer strength is
limited by about a few megahertz due to physical constraints of the
atom energy configuration and laser power \citep{Saskin2019}. We
further calculate the mode decay rates for an $N_{S}=5$ cavity and
presents the results in Fig.~\ref{fig:architecture}(b). A typical
$\kappa\sim10^{-3}\omega_{0}$ implies that the cavity mode can survive
for about a thousand times of momentum exchanges before it vanishes.

\begin{figure}[t]
\begin{centering}
\includegraphics[width=8.5cm]{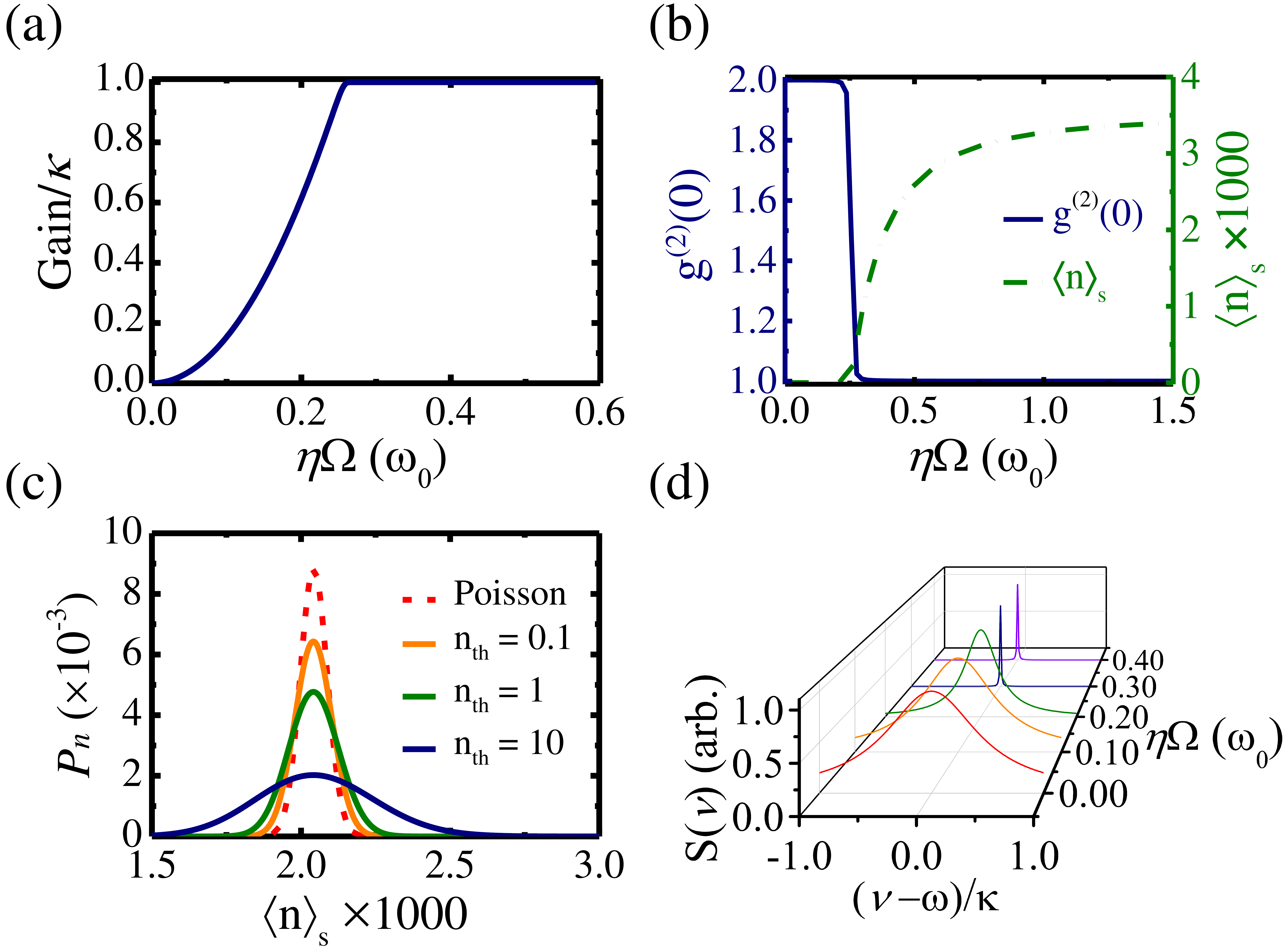}
\par\end{centering}
\caption{(a) Gain as a function of driving strength in terms of $\eta\Omega$
on blue side-band resonance $\delta_{b}=0$. The lasing threshold
is at $\eta\Omega_{c}=0.25\omega_{0}$. (b) Second-order coherence
$g^{(2)}$ (left vertical axis) and the mean phonon number (right
vertical axis) of the cavity mode for varied $\eta\Omega$. (c) Number
distribution for $\eta\Omega=0.4\omega_{0}=2\pi\times0.2$ MHz with
$\langle n\rangle_{s}=2200$. The distribution is broadened by increasing
the noise level. For comparison, the Poisson distribution ($n_{th}=0$)
is plotted in red dashed line. (d) Spectral lineshape for varied $\eta\Omega$
with the peak value normalized to one. In all cases (a)\textendash (d),
we choose $\kappa=6.1\times10^{-3}\omega_{0}=2\pi\times3.1$ kHz given
by $2+2$ tweezers of frequency $2\pi\times2.4$ MHz. The cavity mode
frequency $\omega=2.0\omega_{0}=2\pi\times1.0$ MHz. For (a), (b)
and (d), all the results are under the noise level set by the Doppler
temperature $n_{{\rm th}}=10$ corresponding to the $^{40}$Ca$^{+}$
ion with natural linewidth $\gamma=2\pi\times21.6$ MHz.\label{fig:g2_poisson}}
\end{figure}

\textit{Single-mode phonon lasing.} \textendash{} We now look into
the phonon lasing mechanism of a resonator containing only $N_{S}=1$
ion, which is pumped by lasers resonant with the blue side-band. Typically,
the cavity mode frequency is about hundreds of kilohertz to a few
megahertz, the side-bands are assumed resolvable from the carrier
transition by Raman transitions of a few kilohertz in linewidth. We
describe the evolution of the system by a master equation:
\begin{equation}
\dot{\rho}=-\frac{i}{\hbar}\left[H_{S},\rho\right]-\sum_{\alpha=\pm}\frac{\kappa_{{\rm th}}^{\alpha}}{2}\mathcal{L}_{a}^{\alpha}[\rho]-\frac{\gamma}{2}\mathcal{L}_{\sigma^{-}}^{-}[\rho],\label{eq:mastereq}
\end{equation}
where $\rho$ is the system density matrix governed by the system
Hamiltonian $H_{S}/\hbar=-\delta_{b}\sigma_{z}/2+\eta\Omega\left(a^{\dagger}\sigma^{+}+\sigma^{-}a\right)$
in the rotating frame with $\sigma^{-}=|g\rangle\langle e|$ and $\sigma^{+}=|e\rangle\langle g|$
the atomic lowering and raising operators, respectively, between the
ground state $|g\rangle$ and excited $|e\rangle$ separated by energy
$\hbar\omega_{eg}$; $\sigma_{z}=|e\rangle\langle e|-|g\rangle\langle g|$;
$\gamma$ is the natural linewidth; blue-side band detuning $\delta_{b}\equiv\omega_{L}-\omega_{eg}-\omega$
with driving laser frequency $\omega_{L}$ and cavity mode frequency
$\omega$; $\eta$ is the Lamb-Dicke parameter; $\Omega$ is the Raman
Rabi frequency. The Lindblad superoperators are given by $\mathcal{L}_{a}^{\pm}[\rho]=a^{\pm}a^{\mp}\rho+\rho a^{\pm}a^{\mp}-2a^{\mp}\rho a^{\pm}$
(here we denote $a^{-}=a$ and $a^{+}=a^{\dagger}$ for convenience)
and $\mathcal{L}_{\sigma}^{-}[\rho]=\sigma^{+}\sigma^{-}\rho+\rho\sigma^{+}\sigma^{-}-2\sigma^{-}\rho\sigma^{+}$
with $\kappa_{{\rm th}}^{+}=n_{{\rm th}}\kappa$ and $\kappa_{{\rm th}}^{-}=(n_{{\rm th}}+1)\kappa$,
where the noise level $n_{{\rm th}}$ accounts for nonzero cavity
temperature contribution \citep{Meystre2007} and can be estimated
by $n_{{\rm th}}=[\exp(\hbar\omega/k_{B}T)-1]^{-1}$. Note that the
dynamics of the internal states are much faster than the motional
ones, we can thus assume that the internal degrees of freedom adiabatically
follow the motional operators. The dynamics of the phononic operator
is of the form $\dot{a}=(\mathcal{G}-\kappa)a/2+\text{(noise terms)}$,
which can be obtained by integrating the Heisenberg equations, with
the gain given by
\begin{align}
\mathcal{G}= & \gamma\frac{s}{2}\langle\sigma_{z}\rangle=\sum_{n}\frac{\gamma s}{2(1+ns)}P_{n},\label{eq:gain}
\end{align}
where $s=2\left|\eta\Omega\right|^{2}/[\delta_{b}^{2}+\left(\gamma/2\right)^{2}]$.
We take $\delta_{b}=0$ for simplicity. To determine the population
distribution, we recast the master equation~(\ref{eq:mastereq})
into the rate equations {[}Appendix~\ref{sec:a.rateeq}{]}: 

\begin{align}
\dot{P}_{n}= & -\left[\kappa_{{\rm th}}^{-}n+\kappa_{{\rm th}}^{+}(n+1)\right]P_{n}\nonumber \\
 & +\kappa_{{\rm th}}^{-}(n+1)P_{n+1}+\kappa_{{\rm th}}^{+}nP_{n-1}\label{eq:rateeq}\\
 & -\frac{\gamma s}{2}\left(\frac{n+1}{1+ns}P_{n}-\frac{n}{1+(n-1)s}P_{n-1}\right)\nonumber 
\end{align}
for the probability $P_{n}$ in the motional $n$ state.

Figure~\ref{fig:g2_poisson} shows the gain as a function of pumping
strength in the steady state, where we clearly see the lasing behavior
with $\mathcal{G}/\kappa\rightarrow1$ as the driving strength $\eta\Omega$
overpasses a threshold\textcolor{red}{{} }$\eta\Omega_{c}=0.25\omega_{0}$.
A typical timescale to lasing depends on the steady-state mean phonon
number $\langle n\rangle_{s}$ built. For $\langle n\rangle_{s}\approx2000$,
it is about $14\kappa^{-1}\approx2300\omega_{0}^{-1}$, comparable
to $4.5$ ms. To quantify the degree of lasing, we calculate $\langle n\rangle_{s}$
and second-order coherence $g^{(2)}(0)\equiv\left|\langle a^{\dagger}(0)a^{\dagger}(\tau)a(\tau)a(0)\rangle/\langle a^{\dagger}a\rangle^{2}\right|_{\tau=0}$.
The results are presented in Fig.~\ref{fig:g2_poisson}(b). At low
pumping level below the threshold, $g_{{\rm ph}}^{(2)}(0)$ appears
to be around 2 as the phonon number is small, suggesting a thermal
chaotic phonon state. When $\eta\Omega>\eta\Omega_{c}$, the phonon
number significantly builds up while the $g^{(2)}$ curve abruptly
drops to unity, signaling the emergence of a coherent state, consistent
with the gain profile.

The steady-state phonon number distribution is also obtained from
Eq.~(\ref{eq:rateeq}) and plotted in Fig.~\ref{fig:g2_poisson}(c),
where we includes different profiles for various noise levels. At
zero temperature, the distribution is exactly Poissonian. The rising
noise level gradually broadens the distribution, becoming super-Possonian.
This is however commonly observed in ordinary optical lasers. We also
investigate the line narrowing effect of the lasing mode. The lineshape
is given by {[}Appendix~\ref{sec:a.narrowing}{]}
\begin{align}
S(\nu) & =\frac{\langle n\rangle_{s}}{\left(\nu-\omega\right)^{2}+\Delta\nu^{2}/4},
\end{align}
with $\Delta\nu=\kappa\frac{n_{{\rm th}}}{\langle n\rangle_{s}}+\frac{\gamma}{2\langle n\rangle_{s}}\frac{s}{1+\langle n\rangle_{s}s}$.
One can see clearly from Fig.~\ref{fig:g2_poisson}(d) that the spectral
linewidth becomes narrowed when $\eta\Omega$ exceeds $\eta\Omega_{c}=0.25\omega_{0}$
due to significant increase in the phonon number.

\begin{figure}[t]
\begin{centering}
\includegraphics[width=8.5cm]{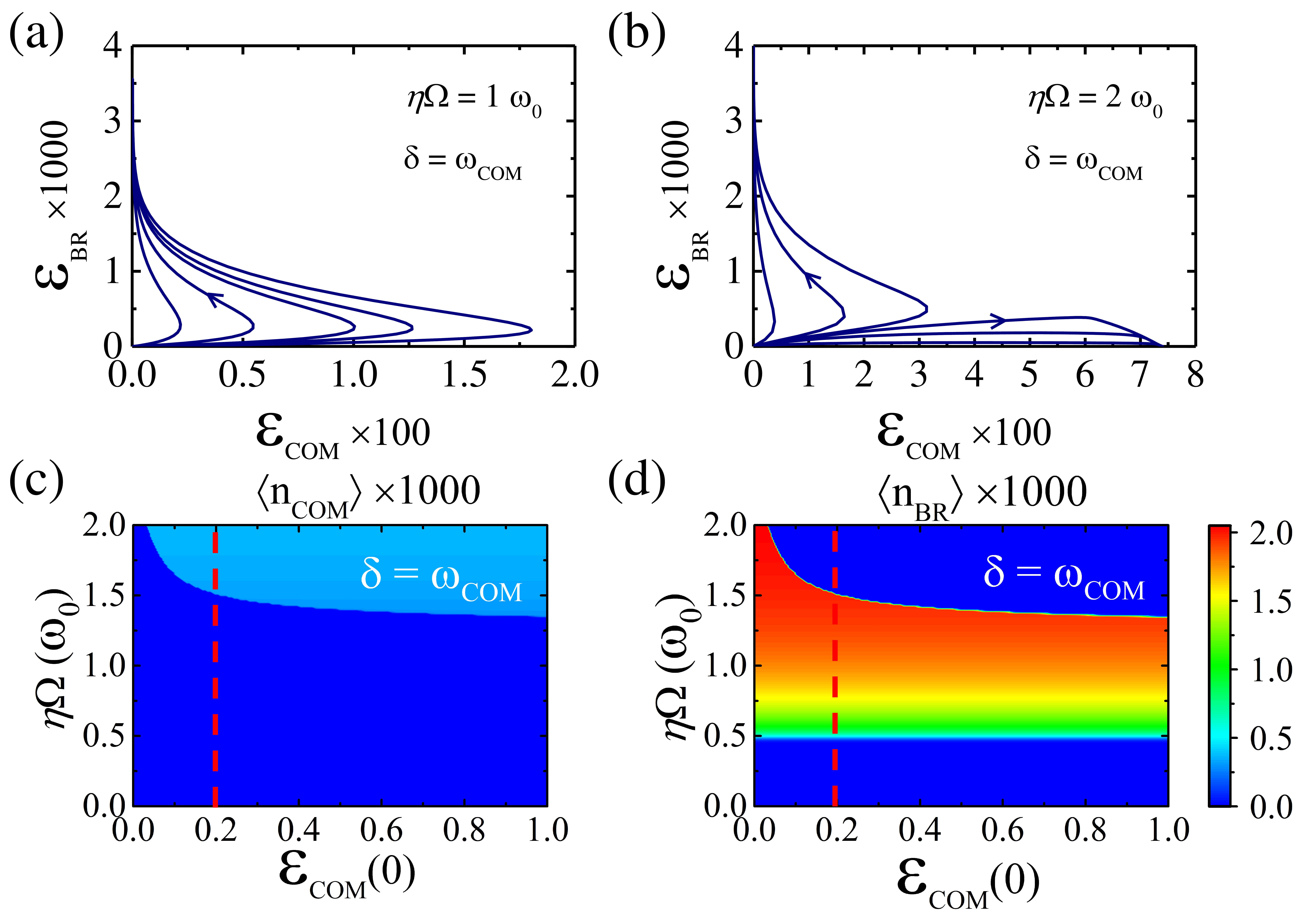}
\par\end{centering}
\caption{Temporal trajectories showing the corresponding classical energy of
the COM and BR modes of the two-mode resonator for (a) $\eta\Omega=1.0\omega_{0}$,
slightly above both thresholds $\eta\Omega_{c}^{{\rm COM}}=0.75\omega_{0}$
and $\eta\Omega_{c}^{{\rm BR}}=0.4\omega_{0}$, and (b) $\eta\Omega=2.0\omega_{0}$,
relatively stronger. In (a), every trajectory approaches to the BR
mode with $\langle n_{{\rm BR}}\rangle_{s}=1780$. In (b), one group
of trajectories approaches to $\langle n_{{\rm BR}}\rangle_{s}=2050$
and the other approaches to $\langle n_{{\rm COM}}\rangle=370$, depending
on the choices of initial distribution of excitation (see text). (c)
Phase diagram in terms of the COM phonon numbers and (d) the BR phonon
numbers for keeping $\mathcal{E}_{\text{BR}}(0)=0.5$.  of both modes.
For the non-lasing mode, the phonon number roughly corresponds to
the thermal level $n_{{\rm th,COM}}=13$ and $n_{{\rm th,BR}}=8.2$.
The line cuts correspond to the profiles discussed in Fig.~\ref{fig:synchronization}.
\label{fig:two_mode}}
\end{figure}

\textit{Two-mode phonon lasing.} \textendash{} We now consider a multi-mode
cavity. Due to growing Hilbert space dimensions and limitation of
computation power, we only focus on calculation for $N_{S}=2$ in
particular, which presents two longitudinal collective modes, the
center-of-mass (COM) mode and the breath (BR) one. We apply on one
of the two ions Raman laser beams to drive the blue-sideband resonances
of the cavity modes. Typically, the frequency difference between the
two modes is about hundreds of kilohertz so it can be assumed optically
resolvable. However, two modes can exchange energy via the atomic
excitation. This can be seen in the gain Eq.~(\ref{eq:gain}), where
its linewidth is comparable to the atomic one $\sim20$ MHz, making
the result insensitive to the laser detuning. Also, due to the increasing
number of coupled rate equations becoming difficult to be managed,
we turn to the Heisenberg equation method by considering the quadrature
operators $X_{q}\equiv a_{q}^{\dagger}+a_{q}$ and $P_{q}\equiv i(a_{q}^{\dagger}-a_{q})$
for $q=\text{COM}$ or $\text{BR}$. We leave the detailed derivation
in Appendix~\ref{sec:a.twomode}, and directly present the calculation
results showing lasing of these modes given various parameters such
as laser detuning, pump power, and starting energy characterized by
initial $\langle X_{j}\rangle$ and $\langle P_{j}\rangle$.

The system dynamics can be characterized by a parameter corresponding
to the classical energy associated with each mode: 
\begin{align}
\mathcal{E}_{q}(t) & \equiv\frac{1}{2}\left[\langle X_{q}(t)\rangle^{2}+\langle P_{q}(t)\rangle^{2}\right],
\end{align}
which is roughly 2 times of the mean phonon number by our convention.
We now demonstrate an exemplary case with $w_{L}+w_{R}=1+1$ tweezers
of frequency $2\pi\times3$ MHz, yielding $\omega_{\text{COM}}=1.6\omega_{0}$
and $\omega_{\text{BR}}=2.5\omega_{0}$ with decay rates $\kappa_{\text{COM}}=0.05\omega_{0}$
and $\kappa_{\text{BR}}=0.01\omega_{0}$, respectively; the thresholds
from the single-mode calculation are given by $\eta\Omega_{c}^{\text{COM}}=0.75\omega_{0}$
and $\eta\Omega_{c}^{\text{BR}}=0.40\omega_{0}$. Note that here we
have set $\eta_{{\rm COM}}=\eta$ so that $\eta_{{\rm BR}}=\sqrt{\omega_{{\rm COM}}/\omega_{{\rm BR}}}\eta$.
We then depict the trajectories of $\mathcal{E}_{j}$ for given lasing
parameters starting from an initial state with specific energies $(\mathcal{E}_{\text{COM}}(0),\mathcal{E}_{\text{BR}}(0))$.

Here we only discuss the cases in particular where the driving strength
surpasses the thresholds of both modes. Our results show that the
two modes cannot sustain lasing simultaneously. The emergence of lasing
in one mode with significant excitation suppresses the other mode
with vanishing phonon numbers. This can be seen in Fig.~\ref{fig:two_mode}(a)
and (b), where the trajectories terminate at one axis point. Explicitly,
Fig.~\ref{fig:two_mode}(a) shows a case where the driving stength
is slightly above both thresholds. When the system starts from low
excitations, both the modes grow in early stages, and evolve into
a steady state in which only one mode survives while the other diminishes.
Since $\eta\Omega_{c}^{\text{BR}}<\eta\Omega_{c}^{\text{COM}}$, it
can be expected that the BR mode reaches lasing first and the COM
mode is suppressed. Note that the final mean phonon number is determined
from competition of the cavity loss and pumping so is independent
from the choice of initial energies. When the pumping strength is
sufficiently strong, at some point the trajectories develop into two
groups corresponding to different lasing modes, as shown in Fig.~\ref{fig:two_mode}(b).
More energy distributed to one mode increases the likeliness of lasing
of that mode.

\begin{figure}[t]
\begin{centering}
\includegraphics[width=7cm]{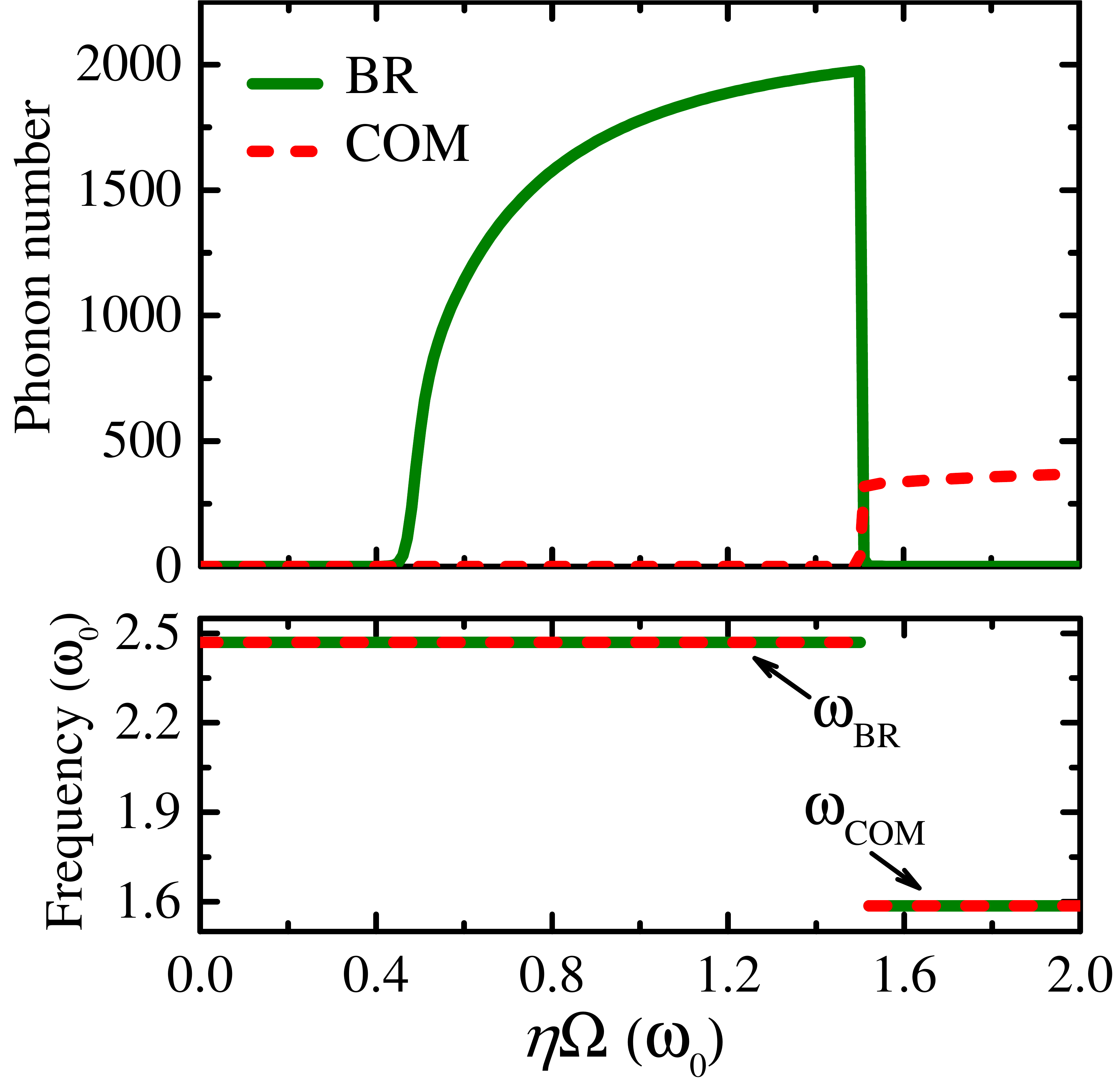}
\par\end{centering}
\caption{(a) The phonon numbers of the COM and BR modes as the driving strength
increases. The first continuous lasing process occurs at the usual
threshold determined by the single-mode calculation. The sharp boundary
emerges at $\eta\Omega=1.5\omega_{0}$. (b) The frequencies of the
two modes. Note that when $\eta\Omega<1.5\omega_{0}$, the $\langle X_{{\rm COM}}(t)\rangle$
(non-lasing mode) follows the frequency of the BR (lasing) mode. Their
roles switch when $\eta\Omega>1.5\omega_{0}$. \label{fig:synchronization}}
\end{figure}

It is helpful to look at the phases defined by the lasing modes in
terms of phonon numbers as shown in Fig.~\ref{fig:two_mode}(c) and
(d). Here we map out the lasing phases by varying the driving strength
and initial energy in the COM mode while keeping $\mathcal{E}_{{\rm BR}}(0)$
constant. From weak to strong pumping strengths, we find that the
system first undergoes a continuous lasing process favoring the BR
mode, suppressing the COM mode, as discussed above. Interestingly,
as pumping increases, up to some point the COM mode suddenly becomes
dominant and prohibits the existence of the BR mode. Fig.~\ref{fig:synchronization}(a)
shows the population change following the line cuts in Fig.~\ref{fig:two_mode}(c)
and (d). The discontinuity of switching from the BR mode to the COM
one signals a first-order phase transition \citep{Sheng2020}. We
further calculate the second-order coherence for the two lasing regimes,
and find that $g^{(2)}(0)$ closely follows the same trend as shown
in Fig.~\ref{fig:g2_poisson}(b) for the continuous lasing process.
Across the sharp boundary, $g^{(2)}(0)$ also switches between 1 and
2 as the phonon number changes.

An interesting effect is that there is a $<1\%$ residual excitation
out of the phonon number of the non-lasing mode. Our calculation shows
that this suppressed mode is forced to catch up with the frequency
of the lasing one by looking at the classical coordinate $\langle X_{q}(t)\rangle$
for $q=\text{COM}$ or $\text{BR}$, as indicated in Fig~\ref{fig:synchronization}(b).
This phenomenon has been studied in the optomechanical two-membrane
system \citep{Sheng2020}, where this effect is visible because the
lasing mode does not dominantly outnumbers the non-lasing one, and
the thermal noise level does not wash the signal out. In our system,
however, we do not expect such synchronization to be detectable because
of the significant thermal contribution.

\textit{Discussion and conclusion.} \textendash{} Finally, we examine
the feasibility and conditions to observe the presented results in
experiments. Take $^{40}$Ca$^{+}$ ions for example. For the single-mode
lasing, we have reached the phonon number $\langle n\rangle_{s}\approx2000$,
which corresponds to a displacement of 0.7 $\mu$m, in a timescale
$2300\omega_{0}^{-1}\sim4.5$ ms. This suggests an ion crystal of
comparable size $\sim2300$ or longer before the reflected wave re-enters
the system. We estimate for $\langle n\rangle_{s}\approx500$ given
by the condition with $1+1$ tweezers of frequency $3$ MHz, the timescale
reduces by an order as $\kappa$ rises, the required size becomes
$<300$. The corresponding displacement is $0.35$ $\mu$m, sufficiently
larger than the thermal contribution $\sim0.05$ $\mu$m.

To sum up, we have presented a tunable prototype of phonon laser based
on an extremely simple architecture analogous to an ordinary optical
resonator. The cavity walls are configurable on demand, controlled
by arrangement of optical tweezers with flexible parameters. As long
as the Markovianity holds, which can be secured by strong enough tweezers,
the lasing mechanism is governed by a master equation, which we have
utilized to calculate the properties including the $g^{(2)}$ coherence,
number distribution, and spectral lineshape. All the results are presented
based on finite temperature calculation set by the Doppler cooling,
suggesting the feasibility in experiments. For the two-mode case,
we have developed the dynamical equations for quadrature operators
of the normal modes. We also demonstrated the lasing mode competition,
and mapped out the phase diagram for the surviving mode. We expect
that the proposed scheme and mathematical methods used apply to more
ion cases, which may yield richer unexplored effects.

\section*{APPENDIX}

\appendix

\section{Decay rates of cavity modes\label{sec:a.kappa}}

In this section we present the detailed calculation of the effective
decay rates for the cavity phonon modes. We divide the system into
the cavity part ($C$) and the environment ($B$) so that the motional
Hamiltonian under the harmonic approximation reads
\begin{align}
H_{m}= & \underbrace{\sum_{i\in C}\frac{p_{i}^{2}}{2m}+\sum_{i,j\in C}A_{ij}z_{i}z_{j}}_{H_{m}^{C}}\nonumber \\
 & +\underbrace{\sum_{i\in B}\frac{p_{i}^{2}}{2m}+\sum_{i,j\in B}A_{ij}z_{i}z_{j}}_{H_{m}^{B}}+\underbrace{\sum_{i\in C,j\in B}A_{ij}z_{i}z_{j}}_{V_{m}^{CB}},\label{eq:Hm_original}
\end{align}
where the elements $A_{ij}$ form the coupling matrix $\mathbf{A}=\mathbf{A}_{C}\oplus\mathbf{A}_{B}+\mathbf{A}_{CB}$,
where $\mathbf{A}_{C}$ and $\mathbf{A}_{B}$ are $N_{C}\times N_{C}$
and $(N-N_{C})\times(N-N_{C})$ submatrices describing the coupling
within the cavity part and environment, respectively; $\oplus$ denotes
the direct sum and $\mathbf{A}_{CB}$ is an $N_{C}\times(N-N_{C})$
matrix containing the interaction between the two subsystems. Each
subsystem's coupling matrix can be diagonalized separately to find
the normal modes represented by the annihilation and creation operator
pair: ($a_{q}$, $a_{q}^{\dagger}$) for the cavity part, where the
mode index $q=1,\cdots,N_{C}$ with $N_{C}=N_{S}+w_{R}+w_{L}$ the
number of ions participating in the cavity and two walls; ($a_{k}$,
$a_{k}^{\dagger}$) for the environment, where the mode index $k=1,\cdots,N-N_{C}$
with $N$ the total number of ions of the entire array. Note that
here we also include the tweezered ions in the subsystem $C$ in order
to retain the smoothness of the dispersion relation of the bath and
secure the Markovianity. However, by doing so, the subsystem $C$
has $w_{L}+w_{R}$ more modes than the supposed $N_{S}$ ones. Fortunately,
those $w_{L}+w_{R}$ modes mainly resulting from tweezered ions are
well separated from the others in frequency, corresponding to spatial
wavevectors very localized on the tweezered sites. This allows us
to identify the rest as the cavity modes with one-to-one correspondence.
In the discussion of phonon lasing, we will only focus on these cavity
modes. 

Under the rotating wave approximation, the Hamiltonian (\ref{eq:Hm_original})
can then be re-written into 
\begin{align}
H_{m}= & \sum_{q=1}^{N_{C}}\hbar\omega_{q}a_{q}^{\dagger}a_{q}+\sum_{k=1}^{N-N_{C}}\hbar\omega_{k}a_{k}^{\dagger}a_{k}\nonumber \\
 & +\sum_{q\in C,k\in B}g_{qk}\left(a_{q}a_{k}^{\dagger}+a_{k}a_{q}^{\dagger}\right),\label{eq:Hm-1}
\end{align}
where $\omega_{q}$ and $\omega_{k}$ are the eigenfrequencies of
the cavity and environment modes, respectively, and $g_{qk}$ deals
with the coupling matrix between mode $q$ in $C$ and mode $k$ in
$B$. Explicitly, $g_{qk}=\frac{\hbar}{2m}\sum_{ij}U_{C,qi}^{T}A_{ij}U_{B,jk}/\sqrt{m^{2}\omega_{q}\omega_{k}}$
also forms an $N_{C}\times(N-N_{C})$ matrix, where $U_{C}$ and $U_{B}$
are the transformation matrices that diagonalize $\mathbf{A}_{C}$
and $\mathbf{A}_{B}$, respectively.

The Heisenberg equations of motion for the field operators then read
\begin{align}
\dot{a}_{q} & =-i\omega_{q}a_{q}-i\sum_{k\in B}g_{qk}a_{k}\\
\dot{a}_{k} & =-i\omega_{k}a_{k}-i\sum_{q\in C}g_{qk}a_{k}.
\end{align}
By integrating out the bath's degrees of freedom, we obtain the following
equation for mode $q\in C$:
\begin{align}
\dot{a}_{q}= & -i\omega_{q}a_{q}-i\sum_{k\in B}g_{qk}a_{k}\left(0\right)e^{-i\omega_{k}t}\nonumber \\
 & -\sum_{k\in B}\left|g_{qk}\right|^{2}a_{q}\int_{0}^{t}dt^{\prime}e^{-i\left(\omega_{k}-\omega_{q}\right)\left(t-t^{\prime}\right)}.
\end{align}
The first term contributes to noise and the second term corresponds
to the decay process, which can be characterized by the rate:
\begin{align}
\kappa_{q} & \approx2\sum_{k\in B}\left|g_{qk}\right|^{2}\int_{0}^{\infty}dt^{\prime}e^{-i\left(\omega_{k}-\omega_{q}\right)t^{\prime}}\approx2\pi\bar{g}_{qq}^{2}\rho_{B}(\omega_{q}).\label{eq:kappa-1}
\end{align}
For a given finite $N$, since the cavity and environment degrees
of freedom are both discrete and these modes may not overlap, we have
calculated the above summation numerically by plugging in the actual
parameters. As long as $N\gg N_{C}$, we find the results have been
found to be consistent with the last approximation, where we have
taken the continuum limit and numerically computed the density of
states $\rho_{B}(\omega)$ for the bath's degrees, and obtained $\bar{g}_{qq}^{2}$
by coarse-graining $|g_{qk}|^{2}$ over a small range of $\omega_{k}\approx\omega_{q}$,
that is, $\bar{g}_{qq}^{2}\equiv\langle|g_{qk}|^{2}\rangle_{\omega_{k}\approx\omega_{q}}.$
This amounts to justification of validity of Markov approximation.
Also, it should be emphasized that here we only focus on longitudinal
modes so that the bath modes constitute a broadband-like spectrum.
The approximation breaks down for transverse modes for its extreme
narrow-band spectral structure.

\begin{figure}[t]
\begin{centering}
\includegraphics[width=8.5cm]{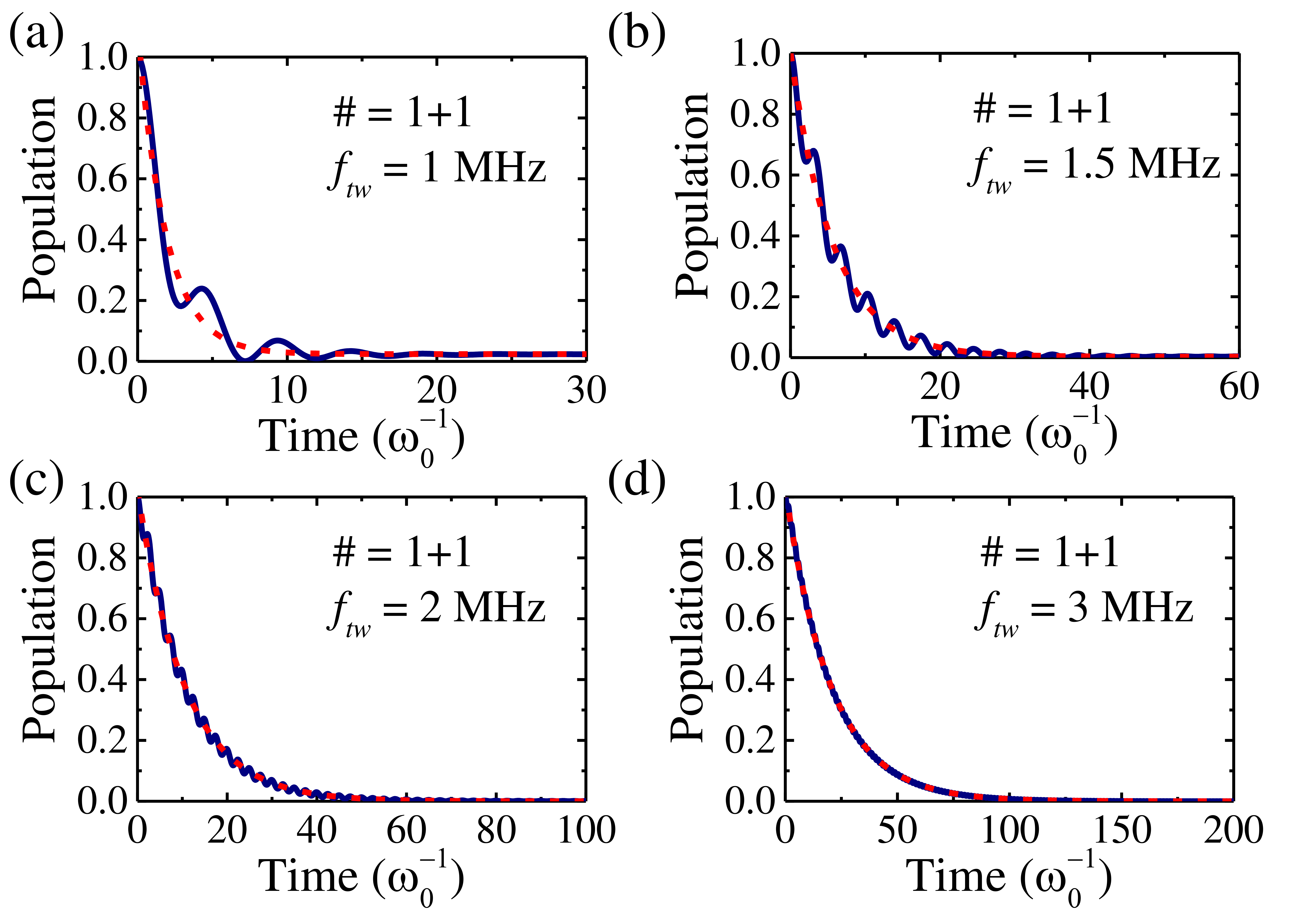}
\par\end{centering}
\caption{We show the real-time population profiles (blue curves) of a single-ion
resonator given varied tweezer frequencies. These curves are fitted
to exponential profiles (red dashed lines) characterized by the decay
rates calculated according to Eq.~(\ref{eq:kappa-1}). These results
are based on calculation considering an ion crystal of more than 1000
$^{40}$Ca$^{+}$ ions with ion separation $7$ $\mu$m. \label{fig:kappa}}
\end{figure}

To verify our calculation, we also look at the real-time population
profile numerically by explicitly including the all motional degrees
of the entire array without approximation. As shown in Fig.~\ref{fig:kappa},
when the tweezer frequency is not strong enough compared to $\omega_{0}$,
we find visible large oscillations causing a certain degree of non-Markovianity.
But the non-Markovianity can be gradually removed when we increase
the tweezer frequency up to approximately an order of magnitude larger
than $\omega_{0}$. In Fig.~\ref{fig:kappa}(d), where $\omega^{ot}/\omega_{0}=5.9$,
we recover the evolution based on Markovian bath assumption.

\noindent 

\section{Probability rate equation and gain\label{sec:a.rateeq}}

In this section, we derive the master equation and the corresponding
probability rate equations considering the phonon cavity being pumped
by blue-sideband lasers. To simplify our discussion, here we only
focus on the single mode case with $N_{S}=1$ ion. Under the Lamb-Dicke
approximation, the Heisenberg equations of motion for the relevant
operators are given by
\begin{align}
\dot{a} & =-\frac{\kappa}{2}a-i\eta\Omega\sigma^{+}\label{eq:da}\\
\dot{\sigma}^{+} & =\left(-i\delta_{b}-\frac{\gamma}{2}\right)\sigma^{+}-i\eta\Omega a\sigma_{z}\label{eq:da+}\\
\dot{\sigma}_{z} & =-\gamma\left(\sigma_{z}+1\right)+2i\eta\Omega\left(\sigma^{-}a-a^{\dagger}\sigma^{+}\right)\label{eq:dz}
\end{align}
where $\sigma^{-}=|g\rangle\langle e|$ and $\sigma^{+}=|e\rangle\langle g|$
are the atomic lowering and raising operators, respectively, between
the ground state $|g\rangle$ and excited $|e\rangle$ separated by
energy $\hbar\omega_{eg}$; $\sigma_{z}=|e\rangle\langle e|-|g\rangle\langle g|$;
$\gamma$ is the natural linewidth; blue side-band detuning $\delta_{b}\equiv\omega_{L}-\omega_{eg}-\omega$
with driving laser frequency $\omega_{L}$ and cavity mode frequency
$\omega$; $\eta$ is the Lamb-Dicke parameter; $\Omega$ is the Raman
Rabi frequency. Note that the dynamics of the internal states are
much faster than the motional ones, we can thus assume that the internal
degrees of freedom adiabatically follow the motional operators. By
taking $\dot{\sigma}^{+}\approx0$, we immediately obtain 
\begin{align}
\sigma^{+}\approx- & \frac{\eta\Omega a\sigma_{z}}{\delta_{b}-i\frac{\gamma}{2}}\label{eq:s+}
\end{align}
and therefore
\begin{align}
\dot{a}= & -\frac{\kappa}{2}a-\frac{1}{2}\frac{\gamma\left|\eta\Omega\right|^{2}}{\delta_{b}^{2}+\left(\frac{\gamma}{2}\right)^{2}}\sigma_{z}a+\text{shift}+\text{noise}.
\end{align}
Note that the noise term must be present in order to assure $a$ a
valid field operator that satisfies $[a,a^{\dagger}]=1$. Both the
shift and noise terms are irrelevant for current discussion. The gain
can now be identified as 
\begin{align}
\mathcal{G} & =-\frac{\gamma\left|\eta\Omega\right|^{2}}{\delta_{b}^{2}+\left(\frac{\gamma}{2}\right)^{2}}\sigma_{z}.\label{eq:gain-1}
\end{align}
Also, by substitution of Eq.~(\ref{eq:s+}) into Eq.~(\ref{eq:dz})
as $\dot{\sigma}_{z}\approx0$, we have
\begin{align}
\sigma_{z} & =-\left(I+\frac{2}{\delta_{b}^{2}+\left(\frac{\gamma}{2}\right)^{2}}\text{\ensuremath{\left|\eta\Omega\right|^{2}}}a^{\dagger}a\right)^{-1}\nonumber \\
 & =-\underset{n}{\sum}\frac{1}{1+ns}\left|n\right\rangle \left\langle n\right|
\end{align}
and
\begin{align}
\eta\Omega\sigma^{+} & =\frac{s}{2}(\delta_{b}+i\gamma)b
\end{align}
where $s=\frac{2\left|\eta\Omega\right|^{2}}{\delta_{b}^{2}+\left(\frac{\gamma}{2}\right)^{2}}$
and $b\equiv\underset{n}{\sum}\frac{\sqrt{n+1}}{1+ns}\left|n\right\rangle \left\langle n+1\right|$.
Plugging the atomic operators back to the master equation Eq.~(3)
in the main text, we finally arrive at
\begin{align}
\dot{P}_{n}= & -\frac{\kappa}{2}\left(n_{{\rm th}}+1\right)\left(2nP_{n}-2\left(n+1\right)P_{n+1}\right)\nonumber \\
 & -\frac{\kappa}{2}n_{{\rm th}}\left(2\left(n+1\right)P_{n}-2nP_{n-1}\right)\nonumber \\
 & -\frac{\gamma s}{2}\left(\frac{n+1}{1+ns}P_{n}-\frac{n}{1+\left(n-1\right)s}P_{n-1}\right).\label{eq:rateeq-1}
\end{align}
Here, we have added the thermal contribution characterized by the
noise level $n_{{\rm th}}$, which can be estimated by $n_{{\rm th}}=[\exp(\hbar\omega/k_{B}T)-1]^{-1}$
\citep{Meystre2007}. In the steady-state, the probability can be
computed

\begin{equation}
P_{n}=P_{0}\prod_{k=1}^{n}\frac{\kappa n_{{\rm th}}+\frac{1}{2}\frac{\gamma s}{1+\left(k-1\right)s}}{\kappa\left(n_{{\rm th}}+1\right)}
\end{equation}
with $P_{0}$ the normalization factor such that $\sum_{n}P_{n}=1$.

\section{Line narrowing\label{sec:a.narrowing}}

To find out the spectral lineshape of the phonon field, we first look
at the mean phonon number equation $d\langle a^{\dagger}a\rangle/dt=\sum_{n}n\dot{P}_{n}$.
By substitution of Eq.~(\ref{eq:rateeq-1}), we obtain \citep{Meystre2007}
\begin{align}
\frac{d}{dt}\left\langle n\right\rangle = & \sum_{n}\Big\{\left(\frac{\gamma}{2}\frac{s}{1+n\cdot s}-\kappa\right)nP_{n}\nonumber \\
 & +\left(\kappa n_{{\rm th}}+\frac{\gamma}{2}\frac{s}{1+ns}\right)P_{n}\Big\}\nonumber \\
\approx & \left(\frac{\gamma}{2}\frac{s}{1+\left\langle n\right\rangle s}-\kappa\right)\left\langle n\right\rangle \nonumber \\
 & +\kappa n_{{\rm th}}+\frac{\gamma}{2}\frac{s}{1+\left\langle n\right\rangle s},\label{eq:dndt}
\end{align}
where we have approximated $n$ in the denominator of the summand
by its instantaneous mean value $\langle n\rangle$. We thus can identify
the gain $\mathcal{G}(t)=\frac{\gamma}{2}\frac{s}{1+\left\langle n\right\rangle s}$,
consistent with Eq.~(\ref{eq:gain-1}) except $n$ is taken to be
the mean value. In the steady state, $\langle n\rangle\rightarrow\langle n\rangle_{s}$,
and
\begin{align}
\mathcal{G} & \rightarrow\kappa\left(1-\frac{n_{{\rm th}}}{\langle n\rangle_{s}}\right)-\frac{\gamma}{2\langle n\rangle_{s}}\frac{s}{1+\langle n\rangle_{s}s}.\label{eq:gain_spectrum}
\end{align}
On the other hand, the Langevin equation reads 
\begin{equation}
\dot{a}\left(t\right)=\left[\mathcal{G}-\left(\frac{\kappa}{2}+i\left(\omega-\nu\right)\right)\right]a\left(t\right)+\text{noise terms},\label{eq:adot}
\end{equation}
where $\nu$ is the probe frequency understood as a Fourier component.
By the quantum regression theorem, we can acquire the spectral lineshape
\begin{align}
S(\nu) & =\mathcal{F}[\langle a^{\dagger}(\tau)a(0)\rangle]\nonumber \\
 & =\frac{\langle n\rangle_{s}}{\left(\nu-\omega\right)^{2}+\Delta\nu^{2}/4},\label{eq:spectrum}
\end{align}
with $\Delta\nu=\kappa\frac{n_{{\rm th}}}{\langle n\rangle_{s}}+\frac{\gamma}{2\langle n\rangle_{s}}\frac{s}{1+\langle n\rangle_{s}s}$,
and $\mathcal{F}$ denotes the Fourier transform.

\section{Model for phonon lasing - two mode\label{sec:a.twomode}}

The rate equation approach summarized in Sec.~\ref{sec:a.rateeq}
also applies to multi-mode cavities. However, the dimension of the
required Hilbert space grows exponentially with the number of modes,
making the calculation intractable even for $N_{S}=2$. Here, we adopt
another method based on the normal-mode quadrature operators \citep{Sheng2020}.

The Hamiltonian that describes the two-mode phonon cavity reads
\begin{align}
H/\hbar & =-\frac{\delta}{2}\sigma_{i}^{z}+\sum_{q}\omega_{q}a_{q}^{\dagger}a_{q}+\sum_{i,q}\frac{\eta_{q}\Omega_{i}}{\sqrt{2}}\left(\sigma_{i}^{+}a_{q}^{\dagger}+a_{q}\sigma_{i}^{-}\right)
\end{align}
where $\delta=\omega_{L}-\omega_{eg}$; $a_{q}$ and $a_{q}^{\dagger}$
are phonon field operators of the $q$th cavity mode; $\sigma_{i}^{-}$
and $\sigma_{i}^{+}$ are atomic operators of the $i$th atom driven
by a laser of frequency $\omega_{L}$ and Rabi frequency $\Omega_{i}$;
$\eta_{q}$ is the corresponding Lamb-Dicke parameter of the $q$th
mode. Now, the normal-mode quadrature operators are defined

\begin{equation}
\begin{cases}
X_{q}=a_{q}^{\dagger}+a_{q}\\
P_{q}=i\left(a_{q}^{\dagger}-a_{q}\right).
\end{cases}
\end{equation}
Then the Langevin equations of the quadrature operators of different
orders are given by 

\begin{widetext}

\begin{align}
\frac{d}{dt}\left(P_{q}^{n}X_{q}^{m}\right)= & -n\omega_{q}\left(P_{q}^{n-1}X_{q}^{m+1}+i\left(n-1\right)P_{q}^{n-2}X_{q}^{m}\right)\nonumber \\
 & +m\omega_{q}\left(P_{q}^{n+1}X_{q}^{m-1}+i\left(m-1\right)P_{q}^{n}X_{q}^{m-2}\right)\nonumber \\
 & -\frac{1}{2}\kappa_{q}nP_{q}^{n}X_{q}^{m}+\frac{1}{2}\kappa_{q}\left(2n_{{\rm th},q}+1\right)\left(n\left(n-1\right)P_{q}^{n-2}X_{q}^{m}\right)\nonumber \\
 & -\frac{1}{2}\kappa_{q}mP_{q}^{n}X_{q}^{m}+\frac{1}{2}\kappa_{q}\left(2n_{{\rm th},q}+1\right)\left(m\left(m-1\right)P_{q}^{n}X_{q}^{m-2}\right)\nonumber \\
 & -n\sum_{i,q}\frac{\eta_{q}\Omega_{i}}{\sqrt{2}}\left(\sigma_{i}^{-}+\sigma_{i}^{+}\right)P_{q}^{n-1}X_{q}^{m}\nonumber \\
 & -im\sum_{i,q}\frac{\eta_{q}\Omega_{i}}{\sqrt{2}}\left(\sigma_{i}^{-}-\sigma_{i}^{+}\right)P_{q}^{n}X_{q}^{m-1},\label{eq:pnxm}\\
\frac{d}{dt}\sigma_{i}^{-}= & \left(i\delta-\frac{\gamma}{2}\right)\sigma_{i}^{-}+i\frac{1}{2}\sum_{q}\frac{\eta_{q}\Omega_{i}}{\sqrt{2}}\sigma_{z,i}\left(X_{q}-iP_{q}\right),\label{eq:s-}\\
\frac{d}{dt}\sigma_{z,i}= & -\gamma\left(\sigma_{z,i}+1\right)-i\sum_{q}\frac{\eta_{q}\Omega_{i}}{\sqrt{2}}\left(\sigma_{z,i}^{+}\left(X_{q}-iP_{q}\right)-\text{H.c.}\right)\label{eq:sz}
\end{align}
\end{widetext}Note that different modes decay independently but are
still coupled through sharing the atomic states, as revealed by the
last terms of Eqs.~(\ref{eq:s-}) and (\ref{eq:sz}). The second-order
coherence can be calculated directly by $g_{q}^{\left(2\right)}\left(0\right)=\left(\left\langle n_{q}^{2}\right\rangle -\left\langle n_{q}\right\rangle \right)/\left\langle n_{q}\right\rangle ^{2}$,
where
\begin{align}
\left\langle n_{q}\right\rangle = & \frac{1}{4}\left(\left\langle X_{q}^{2}\right\rangle +\left\langle P_{q}^{2}\right\rangle \right)-\frac{1}{2}
\end{align}
and

\begin{align}
\left\langle n_{q}^{2}\right\rangle = & \frac{1}{16}\left(\left\langle X_{q}^{4}\right\rangle +\left\langle P_{q}^{4}\right\rangle +2Re\left\langle P_{q}^{2}X_{q}^{2}\right\rangle \right)\nonumber \\
 & -\frac{1}{4}\left(\left\langle X_{q}^{2}\right\rangle +\left\langle P_{q}^{2}\right\rangle +2Im\left\langle P_{q}X_{q}\right\rangle +1\right).
\end{align}

\section*{Acknowledgments}

We thank the support from MOST of Taiwan under Grant No. 109-2112-M-002-022
and National Taiwan University under Grant No. NTU-CC-109L892006.
GDL thanks Ming-Shien Chang and Mishkat Bhattacharya for valuable
discussion and feedback.

\bibliographystyle{apsrev4-2}
\bibliography{phononlaser_ref}

\begin{thebibliography}{24}%
\makeatletter
\providecommand \@ifxundefined [1]{%
 \@ifx{#1\undefined}
}%
\providecommand \@ifnum [1]{%
 \ifnum #1\expandafter \@firstoftwo
 \else \expandafter \@secondoftwo
 \fi
}%
\providecommand \@ifx [1]{%
 \ifx #1\expandafter \@firstoftwo
 \else \expandafter \@secondoftwo
 \fi
}%
\providecommand \natexlab [1]{#1}%
\providecommand \enquote  [1]{``#1''}%
\providecommand \bibnamefont  [1]{#1}%
\providecommand \bibfnamefont [1]{#1}%
\providecommand \citenamefont [1]{#1}%
\providecommand \href@noop [0]{\@secondoftwo}%
\providecommand \href [0]{\begingroup \@sanitize@url \@href}%
\providecommand \@href[1]{\@@startlink{#1}\@@href}%
\providecommand \@@href[1]{\endgroup#1\@@endlink}%
\providecommand \@sanitize@url [0]{\catcode `\\12\catcode `\$12\catcode
  `\&12\catcode `\#12\catcode `\^12\catcode `\_12\catcode `\%12\relax}%
\providecommand \@@startlink[1]{}%
\providecommand \@@endlink[0]{}%
\providecommand \url  [0]{\begingroup\@sanitize@url \@url }%
\providecommand \@url [1]{\endgroup\@href {#1}{\urlprefix }}%
\providecommand \urlprefix  [0]{URL }%
\providecommand \Eprint [0]{\href }%
\providecommand \doibase [0]{https://doi.org/}%
\providecommand \selectlanguage [0]{\@gobble}%
\providecommand \bibinfo  [0]{\@secondoftwo}%
\providecommand \bibfield  [0]{\@secondoftwo}%
\providecommand \translation [1]{[#1]}%
\providecommand \BibitemOpen [0]{}%
\providecommand \bibitemStop [0]{}%
\providecommand \bibitemNoStop [0]{.\EOS\space}%
\providecommand \EOS [0]{\spacefactor3000\relax}%
\providecommand \BibitemShut  [1]{\csname bibitem#1\endcsname}%
\let\auto@bib@innerbib\@empty
\bibitem [{\citenamefont {Cirac}\ and\ \citenamefont
  {Zoller}(1995)}]{Cirac1995}%
  \BibitemOpen
  \bibfield  {author} {\bibinfo {author} {\bibfnamefont {J.~I.}\ \bibnamefont
  {Cirac}}\ and\ \bibinfo {author} {\bibfnamefont {P.}~\bibnamefont {Zoller}},\
  }\href {https://doi.org/10.1103/PhysRevLett.74.4091} {\bibfield  {journal}
  {\bibinfo  {journal} {Phys. Rev. Lett.}\ }\textbf {\bibinfo {volume} {74}},\
  \bibinfo {pages} {4091} (\bibinfo {year} {1995})}\BibitemShut {NoStop}%
\bibitem [{\citenamefont {Vahala}\ \emph {et~al.}(2009)\citenamefont {Vahala},
  \citenamefont {Herrmann}, \citenamefont {Kn{\"u}nz}, \citenamefont
  {Batteiger}, \citenamefont {Saathoff}, \citenamefont {H{\"a}nsch},\ and\
  \citenamefont {Udem}}]{Vahala2009}%
  \BibitemOpen
  \bibfield  {author} {\bibinfo {author} {\bibfnamefont {K.}~\bibnamefont
  {Vahala}}, \bibinfo {author} {\bibfnamefont {M.}~\bibnamefont {Herrmann}},
  \bibinfo {author} {\bibfnamefont {S.}~\bibnamefont {Kn{\"u}nz}}, \bibinfo
  {author} {\bibfnamefont {V.}~\bibnamefont {Batteiger}}, \bibinfo {author}
  {\bibfnamefont {G.}~\bibnamefont {Saathoff}}, \bibinfo {author}
  {\bibfnamefont {T.~W.}\ \bibnamefont {H{\"a}nsch}},\ and\ \bibinfo {author}
  {\bibfnamefont {T.}~\bibnamefont {Udem}},\ }\href
  {https://doi.org/10.1038/nphys1367} {\bibfield  {journal} {\bibinfo
  {journal} {Nature Physics}\ }\textbf {\bibinfo {volume} {5}},\ \bibinfo
  {pages} {682} (\bibinfo {year} {2009})}\BibitemShut {NoStop}%
\bibitem [{\citenamefont {Kn\"unz}\ \emph {et~al.}(2010)\citenamefont
  {Kn\"unz}, \citenamefont {Herrmann}, \citenamefont {Batteiger}, \citenamefont
  {Saathoff}, \citenamefont {H\"ansch}, \citenamefont {Vahala},\ and\
  \citenamefont {Udem}}]{Knunz2010}%
  \BibitemOpen
  \bibfield  {author} {\bibinfo {author} {\bibfnamefont {S.}~\bibnamefont
  {Kn\"unz}}, \bibinfo {author} {\bibfnamefont {M.}~\bibnamefont {Herrmann}},
  \bibinfo {author} {\bibfnamefont {V.}~\bibnamefont {Batteiger}}, \bibinfo
  {author} {\bibfnamefont {G.}~\bibnamefont {Saathoff}}, \bibinfo {author}
  {\bibfnamefont {T.~W.}\ \bibnamefont {H\"ansch}}, \bibinfo {author}
  {\bibfnamefont {K.}~\bibnamefont {Vahala}},\ and\ \bibinfo {author}
  {\bibfnamefont {T.}~\bibnamefont {Udem}},\ }\href
  {https://doi.org/10.1103/PhysRevLett.105.013004} {\bibfield  {journal}
  {\bibinfo  {journal} {Phys. Rev. Lett.}\ }\textbf {\bibinfo {volume} {105}},\
  \bibinfo {pages} {013004} (\bibinfo {year} {2010})}\BibitemShut {NoStop}%
\bibitem [{\citenamefont {Ip}\ \emph {et~al.}(2018)\citenamefont {Ip},
  \citenamefont {Ransford}, \citenamefont {Jayich}, \citenamefont {Long},
  \citenamefont {Roman},\ and\ \citenamefont {Campbell}}]{Ip2018}%
  \BibitemOpen
  \bibfield  {author} {\bibinfo {author} {\bibfnamefont {M.}~\bibnamefont
  {Ip}}, \bibinfo {author} {\bibfnamefont {A.}~\bibnamefont {Ransford}},
  \bibinfo {author} {\bibfnamefont {A.~M.}\ \bibnamefont {Jayich}}, \bibinfo
  {author} {\bibfnamefont {X.}~\bibnamefont {Long}}, \bibinfo {author}
  {\bibfnamefont {C.}~\bibnamefont {Roman}},\ and\ \bibinfo {author}
  {\bibfnamefont {W.~C.}\ \bibnamefont {Campbell}},\ }\href
  {https://doi.org/10.1103/PhysRevLett.121.043201} {\bibfield  {journal}
  {\bibinfo  {journal} {Phys. Rev. Lett.}\ }\textbf {\bibinfo {volume} {121}},\
  \bibinfo {pages} {043201} (\bibinfo {year} {2018})}\BibitemShut {NoStop}%
\bibitem [{\citenamefont {Kabuss}\ \emph {et~al.}(2012)\citenamefont {Kabuss},
  \citenamefont {Carmele}, \citenamefont {Brandes},\ and\ \citenamefont
  {Knorr}}]{Kabuss2012}%
  \BibitemOpen
  \bibfield  {author} {\bibinfo {author} {\bibfnamefont {J.}~\bibnamefont
  {Kabuss}}, \bibinfo {author} {\bibfnamefont {A.}~\bibnamefont {Carmele}},
  \bibinfo {author} {\bibfnamefont {T.}~\bibnamefont {Brandes}},\ and\ \bibinfo
  {author} {\bibfnamefont {A.}~\bibnamefont {Knorr}},\ }\href
  {https://doi.org/10.1103/PhysRevLett.109.054301} {\bibfield  {journal}
  {\bibinfo  {journal} {Phys. Rev. Lett.}\ }\textbf {\bibinfo {volume} {109}},\
  \bibinfo {pages} {054301} (\bibinfo {year} {2012})}\BibitemShut {NoStop}%
\bibitem [{\citenamefont {Kabuss}\ \emph {et~al.}(2013)\citenamefont {Kabuss},
  \citenamefont {Carmele},\ and\ \citenamefont {Knorr}}]{Kabuss2013}%
  \BibitemOpen
  \bibfield  {author} {\bibinfo {author} {\bibfnamefont {J.}~\bibnamefont
  {Kabuss}}, \bibinfo {author} {\bibfnamefont {A.}~\bibnamefont {Carmele}},\
  and\ \bibinfo {author} {\bibfnamefont {A.}~\bibnamefont {Knorr}},\ }\href
  {https://doi.org/10.1103/PhysRevB.88.064305} {\bibfield  {journal} {\bibinfo
  {journal} {Phys. Rev. B}\ }\textbf {\bibinfo {volume} {88}},\ \bibinfo
  {pages} {064305} (\bibinfo {year} {2013})}\BibitemShut {NoStop}%
\bibitem [{\citenamefont {Khaetskii}\ \emph {et~al.}(2013)\citenamefont
  {Khaetskii}, \citenamefont {Golovach}, \citenamefont {Hu},\ and\
  \citenamefont {\v{Z}uti\'{c}}}]{Khaetskii2013}%
  \BibitemOpen
  \bibfield  {author} {\bibinfo {author} {\bibfnamefont {A.}~\bibnamefont
  {Khaetskii}}, \bibinfo {author} {\bibfnamefont {V.~N.}\ \bibnamefont
  {Golovach}}, \bibinfo {author} {\bibfnamefont {X.}~\bibnamefont {Hu}},\ and\
  \bibinfo {author} {\bibfnamefont {I.}~\bibnamefont {\v{Z}uti\'{c}}},\ }\href
  {https://doi.org/10.1103/PhysRevLett.111.186601} {\bibfield  {journal}
  {\bibinfo  {journal} {Phys. Rev. Lett.}\ }\textbf {\bibinfo {volume} {111}},\
  \bibinfo {pages} {186601} (\bibinfo {year} {2013})}\BibitemShut {NoStop}%
\bibitem [{\citenamefont {Grudinin}\ \emph {et~al.}(2010)\citenamefont
  {Grudinin}, \citenamefont {Lee}, \citenamefont {Painter},\ and\ \citenamefont
  {Vahala}}]{Grudinin2010}%
  \BibitemOpen
  \bibfield  {author} {\bibinfo {author} {\bibfnamefont {I.~S.}\ \bibnamefont
  {Grudinin}}, \bibinfo {author} {\bibfnamefont {H.}~\bibnamefont {Lee}},
  \bibinfo {author} {\bibfnamefont {O.}~\bibnamefont {Painter}},\ and\ \bibinfo
  {author} {\bibfnamefont {K.~J.}\ \bibnamefont {Vahala}},\ }\href
  {https://doi.org/10.1103/PhysRevLett.104.083901} {\bibfield  {journal}
  {\bibinfo  {journal} {Phys. Rev. Lett.}\ }\textbf {\bibinfo {volume} {104}},\
  \bibinfo {pages} {083901} (\bibinfo {year} {2010})}\BibitemShut {NoStop}%
\bibitem [{\citenamefont {Beardsley}\ \emph {et~al.}(2010)\citenamefont
  {Beardsley}, \citenamefont {Akimov}, \citenamefont {Henini},\ and\
  \citenamefont {Kent}}]{Beardsley2010}%
  \BibitemOpen
  \bibfield  {author} {\bibinfo {author} {\bibfnamefont {R.~P.}\ \bibnamefont
  {Beardsley}}, \bibinfo {author} {\bibfnamefont {A.~V.}\ \bibnamefont
  {Akimov}}, \bibinfo {author} {\bibfnamefont {M.}~\bibnamefont {Henini}},\
  and\ \bibinfo {author} {\bibfnamefont {A.~J.}\ \bibnamefont {Kent}},\ }\href
  {https://doi.org/10.1103/PhysRevLett.104.085501} {\bibfield  {journal}
  {\bibinfo  {journal} {Phys. Rev. Lett.}\ }\textbf {\bibinfo {volume} {104}},\
  \bibinfo {pages} {085501} (\bibinfo {year} {2010})}\BibitemShut {NoStop}%
\bibitem [{\citenamefont {Mahboob}\ \emph {et~al.}(2013)\citenamefont
  {Mahboob}, \citenamefont {Nishiguchi}, \citenamefont {Fujiwara},\ and\
  \citenamefont {Yamaguchi}}]{Mahboob2013}%
  \BibitemOpen
  \bibfield  {author} {\bibinfo {author} {\bibfnamefont {I.}~\bibnamefont
  {Mahboob}}, \bibinfo {author} {\bibfnamefont {K.}~\bibnamefont {Nishiguchi}},
  \bibinfo {author} {\bibfnamefont {A.}~\bibnamefont {Fujiwara}},\ and\
  \bibinfo {author} {\bibfnamefont {H.}~\bibnamefont {Yamaguchi}},\ }\href
  {https://doi.org/10.1103/PhysRevLett.110.127202} {\bibfield  {journal}
  {\bibinfo  {journal} {Phys. Rev. Lett.}\ }\textbf {\bibinfo {volume} {110}},\
  \bibinfo {pages} {127202} (\bibinfo {year} {2013})}\BibitemShut {NoStop}%
\bibitem [{\citenamefont {Kemiktarak}\ \emph {et~al.}(2014)\citenamefont
  {Kemiktarak}, \citenamefont {Durand}, \citenamefont {Metcalfe},\ and\
  \citenamefont {Lawall}}]{Kemiktarak2014}%
  \BibitemOpen
  \bibfield  {author} {\bibinfo {author} {\bibfnamefont {U.}~\bibnamefont
  {Kemiktarak}}, \bibinfo {author} {\bibfnamefont {M.}~\bibnamefont {Durand}},
  \bibinfo {author} {\bibfnamefont {M.}~\bibnamefont {Metcalfe}},\ and\
  \bibinfo {author} {\bibfnamefont {J.}~\bibnamefont {Lawall}},\ }\href
  {https://doi.org/10.1103/PhysRevLett.113.030802} {\bibfield  {journal}
  {\bibinfo  {journal} {Phys. Rev. Lett.}\ }\textbf {\bibinfo {volume} {113}},\
  \bibinfo {pages} {030802} (\bibinfo {year} {2014})}\BibitemShut {NoStop}%
\bibitem [{\citenamefont {Jing}\ \emph {et~al.}(2014)\citenamefont {Jing},
  \citenamefont {\"Ozdemir}, \citenamefont {L\"u}, \citenamefont {Zhang},
  \citenamefont {Yang},\ and\ \citenamefont {Nori}}]{Jing2014}%
  \BibitemOpen
  \bibfield  {author} {\bibinfo {author} {\bibfnamefont {H.}~\bibnamefont
  {Jing}}, \bibinfo {author} {\bibfnamefont {S.~K.}\ \bibnamefont {\"Ozdemir}},
  \bibinfo {author} {\bibfnamefont {X.-Y.}\ \bibnamefont {L\"u}}, \bibinfo
  {author} {\bibfnamefont {J.}~\bibnamefont {Zhang}}, \bibinfo {author}
  {\bibfnamefont {L.}~\bibnamefont {Yang}},\ and\ \bibinfo {author}
  {\bibfnamefont {F.}~\bibnamefont {Nori}},\ }\href
  {https://doi.org/10.1103/PhysRevLett.113.053604} {\bibfield  {journal}
  {\bibinfo  {journal} {Phys. Rev. Lett.}\ }\textbf {\bibinfo {volume} {113}},\
  \bibinfo {pages} {053604} (\bibinfo {year} {2014})}\BibitemShut {NoStop}%
\bibitem [{\citenamefont {Zhang}\ \emph {et~al.}(2018)\citenamefont {Zhang},
  \citenamefont {Peng}, \citenamefont {{\"O}zdemir}, \citenamefont {Pichler},
  \citenamefont {Krimer}, \citenamefont {Zhao}, \citenamefont {Nori},
  \citenamefont {Liu}, \citenamefont {Rotter},\ and\ \citenamefont
  {Yang}}]{Zhang2018}%
  \BibitemOpen
  \bibfield  {author} {\bibinfo {author} {\bibfnamefont {J.}~\bibnamefont
  {Zhang}}, \bibinfo {author} {\bibfnamefont {B.}~\bibnamefont {Peng}},
  \bibinfo {author} {\bibfnamefont {{\c{S}}.~K.}\ \bibnamefont {{\"O}zdemir}},
  \bibinfo {author} {\bibfnamefont {K.}~\bibnamefont {Pichler}}, \bibinfo
  {author} {\bibfnamefont {D.~O.}\ \bibnamefont {Krimer}}, \bibinfo {author}
  {\bibfnamefont {G.}~\bibnamefont {Zhao}}, \bibinfo {author} {\bibfnamefont
  {F.}~\bibnamefont {Nori}}, \bibinfo {author} {\bibfnamefont {Y.-x.}\
  \bibnamefont {Liu}}, \bibinfo {author} {\bibfnamefont {S.}~\bibnamefont
  {Rotter}},\ and\ \bibinfo {author} {\bibfnamefont {L.}~\bibnamefont {Yang}},\
  }\href {https://doi.org/10.1038/s41566-018-0213-5} {\bibfield  {journal}
  {\bibinfo  {journal} {Nature Photonics}\ }\textbf {\bibinfo {volume} {12}},\
  \bibinfo {pages} {479} (\bibinfo {year} {2018})}\BibitemShut {NoStop}%
\bibitem [{\citenamefont {Jiang}\ \emph {et~al.}(2018)\citenamefont {Jiang},
  \citenamefont {Maayani}, \citenamefont {Carmon}, \citenamefont {Nori},\ and\
  \citenamefont {Jing}}]{Jiang2018}%
  \BibitemOpen
  \bibfield  {author} {\bibinfo {author} {\bibfnamefont {Y.}~\bibnamefont
  {Jiang}}, \bibinfo {author} {\bibfnamefont {S.}~\bibnamefont {Maayani}},
  \bibinfo {author} {\bibfnamefont {T.}~\bibnamefont {Carmon}}, \bibinfo
  {author} {\bibfnamefont {F.}~\bibnamefont {Nori}},\ and\ \bibinfo {author}
  {\bibfnamefont {H.}~\bibnamefont {Jing}},\ }\href
  {https://doi.org/10.1103/PhysRevApplied.10.064037} {\bibfield  {journal}
  {\bibinfo  {journal} {Phys. Rev. Applied}\ }\textbf {\bibinfo {volume}
  {10}},\ \bibinfo {pages} {064037} (\bibinfo {year} {2018})}\BibitemShut
  {NoStop}%
\bibitem [{\citenamefont {Pettit}\ \emph {et~al.}(2019)\citenamefont {Pettit},
  \citenamefont {Ge}, \citenamefont {Kumar}, \citenamefont {Luntz-Martin},
  \citenamefont {Schultz}, \citenamefont {Neukirch}, \citenamefont
  {Bhattacharya},\ and\ \citenamefont {Vamivakas}}]{Pettit2019}%
  \BibitemOpen
  \bibfield  {author} {\bibinfo {author} {\bibfnamefont {R.~M.}\ \bibnamefont
  {Pettit}}, \bibinfo {author} {\bibfnamefont {W.}~\bibnamefont {Ge}}, \bibinfo
  {author} {\bibfnamefont {P.}~\bibnamefont {Kumar}}, \bibinfo {author}
  {\bibfnamefont {D.~R.}\ \bibnamefont {Luntz-Martin}}, \bibinfo {author}
  {\bibfnamefont {J.~T.}\ \bibnamefont {Schultz}}, \bibinfo {author}
  {\bibfnamefont {L.~P.}\ \bibnamefont {Neukirch}}, \bibinfo {author}
  {\bibfnamefont {M.}~\bibnamefont {Bhattacharya}},\ and\ \bibinfo {author}
  {\bibfnamefont {A.~N.}\ \bibnamefont {Vamivakas}},\ }\href
  {https://doi.org/10.1038/s41566-019-0395-5} {\bibfield  {journal} {\bibinfo
  {journal} {Nature Photonics}\ }\textbf {\bibinfo {volume} {13}},\ \bibinfo
  {pages} {402} (\bibinfo {year} {2019})}\BibitemShut {NoStop}%
\bibitem [{\citenamefont {Sheng}\ \emph {et~al.}(2020)\citenamefont {Sheng},
  \citenamefont {Wei}, \citenamefont {Yang},\ and\ \citenamefont
  {Wu}}]{Sheng2020}%
  \BibitemOpen
  \bibfield  {author} {\bibinfo {author} {\bibfnamefont {J.}~\bibnamefont
  {Sheng}}, \bibinfo {author} {\bibfnamefont {X.}~\bibnamefont {Wei}}, \bibinfo
  {author} {\bibfnamefont {C.}~\bibnamefont {Yang}},\ and\ \bibinfo {author}
  {\bibfnamefont {H.}~\bibnamefont {Wu}},\ }\href
  {https://doi.org/10.1103/PhysRevLett.124.053604} {\bibfield  {journal}
  {\bibinfo  {journal} {Phys. Rev. Lett.}\ }\textbf {\bibinfo {volume} {124}},\
  \bibinfo {pages} {053604} (\bibinfo {year} {2020})}\BibitemShut {NoStop}%
\bibitem [{\citenamefont {Khurgin}\ \emph {et~al.}(2012)\citenamefont
  {Khurgin}, \citenamefont {Pruessner}, \citenamefont {Stievater},\ and\
  \citenamefont {Rabinovich}}]{Khurgin2012}%
  \BibitemOpen
  \bibfield  {author} {\bibinfo {author} {\bibfnamefont {J.~B.}\ \bibnamefont
  {Khurgin}}, \bibinfo {author} {\bibfnamefont {M.~W.}\ \bibnamefont
  {Pruessner}}, \bibinfo {author} {\bibfnamefont {T.~H.}\ \bibnamefont
  {Stievater}},\ and\ \bibinfo {author} {\bibfnamefont {W.~S.}\ \bibnamefont
  {Rabinovich}},\ }\href {https://doi.org/10.1103/PhysRevLett.108.223904}
  {\bibfield  {journal} {\bibinfo  {journal} {Phys. Rev. Lett.}\ }\textbf
  {\bibinfo {volume} {108}},\ \bibinfo {pages} {223904} (\bibinfo {year}
  {2012})}\BibitemShut {NoStop}%
\bibitem [{\citenamefont {Shen}\ and\ \citenamefont {Lin}(2020)}]{Shen2020}%
  \BibitemOpen
  \bibfield  {author} {\bibinfo {author} {\bibfnamefont {Y.-C.}\ \bibnamefont
  {Shen}}\ and\ \bibinfo {author} {\bibfnamefont {G.-D.}\ \bibnamefont {Lin}},\
  }\href {https://doi.org/10.1088/1367-2630/ab84b6} {\bibfield  {journal}
  {\bibinfo  {journal} {New Journal of Physics}\ }\textbf {\bibinfo {volume}
  {22}},\ \bibinfo {pages} {053032} (\bibinfo {year} {2020})}\BibitemShut
  {NoStop}%
\bibitem [{\citenamefont {Cirac}\ and\ \citenamefont
  {Zoller}(2000)}]{Cirac2000}%
  \BibitemOpen
  \bibfield  {author} {\bibinfo {author} {\bibfnamefont {J.~I.}\ \bibnamefont
  {Cirac}}\ and\ \bibinfo {author} {\bibfnamefont {P.}~\bibnamefont {Zoller}},\
  }\href {https://doi.org/10.1038/35007021} {\bibfield  {journal} {\bibinfo
  {journal} {Nature}\ }\textbf {\bibinfo {volume} {404}},\ \bibinfo {pages}
  {579} (\bibinfo {year} {2000})}\BibitemShut {NoStop}%
\bibitem [{\citenamefont {Ratcliffe}\ \emph {et~al.}(2018)\citenamefont
  {Ratcliffe}, \citenamefont {Taylor}, \citenamefont {Hope},\ and\
  \citenamefont {Carvalho}}]{Ratcliffe2018}%
  \BibitemOpen
  \bibfield  {author} {\bibinfo {author} {\bibfnamefont {A.~K.}\ \bibnamefont
  {Ratcliffe}}, \bibinfo {author} {\bibfnamefont {R.~L.}\ \bibnamefont
  {Taylor}}, \bibinfo {author} {\bibfnamefont {J.~J.}\ \bibnamefont {Hope}},\
  and\ \bibinfo {author} {\bibfnamefont {A.~R.}\ \bibnamefont {Carvalho}},\
  }\href {https://doi.org/10.1103/PhysRevLett.120.220501} {\bibfield  {journal}
  {\bibinfo  {journal} {Physical Review Letters}\ }\textbf {\bibinfo {volume}
  {120}},\ \bibinfo {pages} {1} (\bibinfo {year} {2018})},\ \Eprint
  {https://arxiv.org/abs/1711.05875} {arXiv:1711.05875} \BibitemShut {NoStop}%
\bibitem [{\citenamefont {Jain}\ \emph {et~al.}(2020)\citenamefont {Jain},
  \citenamefont {Alonso}, \citenamefont {Grau},\ and\ \citenamefont
  {Home}}]{Jain2020}%
  \BibitemOpen
  \bibfield  {author} {\bibinfo {author} {\bibfnamefont {S.}~\bibnamefont
  {Jain}}, \bibinfo {author} {\bibfnamefont {J.}~\bibnamefont {Alonso}},
  \bibinfo {author} {\bibfnamefont {M.}~\bibnamefont {Grau}},\ and\ \bibinfo
  {author} {\bibfnamefont {J.}~\bibnamefont {Home}},\ }\bibfield  {journal}
  {\bibinfo  {journal} {Physical Review X}\ }\textbf {\bibinfo {volume} {10}},\
  \href {https://doi.org/10.1103/physrevx.10.031027}
  {10.1103/physrevx.10.031027} (\bibinfo {year} {2020})\BibitemShut {NoStop}%
\bibitem [{\citenamefont {Zhu}\ \emph {et~al.}(2006)\citenamefont {Zhu},
  \citenamefont {Monroe},\ and\ \citenamefont {Duan}}]{Zhu2006}%
  \BibitemOpen
  \bibfield  {author} {\bibinfo {author} {\bibfnamefont {S.-L.}\ \bibnamefont
  {Zhu}}, \bibinfo {author} {\bibfnamefont {C.}~\bibnamefont {Monroe}},\ and\
  \bibinfo {author} {\bibfnamefont {L.-M.}\ \bibnamefont {Duan}},\ }\href
  {https://doi.org/10.1103/PhysRevLett.97.050505} {\bibfield  {journal}
  {\bibinfo  {journal} {Phys. Rev. Lett.}\ }\textbf {\bibinfo {volume} {97}},\
  \bibinfo {pages} {050505} (\bibinfo {year} {2006})}\BibitemShut {NoStop}%
\bibitem [{\citenamefont {Saskin}\ \emph {et~al.}(2019)\citenamefont {Saskin},
  \citenamefont {Wilson}, \citenamefont {Grinkemeyer},\ and\ \citenamefont
  {Thompson}}]{Saskin2019}%
  \BibitemOpen
  \bibfield  {author} {\bibinfo {author} {\bibfnamefont {S.}~\bibnamefont
  {Saskin}}, \bibinfo {author} {\bibfnamefont {J.}~\bibnamefont {Wilson}},
  \bibinfo {author} {\bibfnamefont {B.}~\bibnamefont {Grinkemeyer}},\ and\
  \bibinfo {author} {\bibfnamefont {J.}~\bibnamefont {Thompson}},\ }\bibfield
  {journal} {\bibinfo  {journal} {Physical Review Letters}\ }\textbf {\bibinfo
  {volume} {122}},\ \href {https://doi.org/10.1103/physrevlett.122.143002}
  {10.1103/physrevlett.122.143002} (\bibinfo {year} {2019})\BibitemShut
  {NoStop}%
\bibitem [{\citenamefont {Pierre~Meystre}(2007)}]{Meystre2007}%
  \BibitemOpen
  \bibfield  {author} {\bibinfo {author} {\bibfnamefont {M.~S.}\ \bibnamefont
  {Pierre~Meystre}},\ }\href@noop {} {\emph {\bibinfo {title} {Elements of
  Quantum Optics}}}\ (\bibinfo  {publisher} {Springer-Verlag Berlin
  Heidelberg},\ \bibinfo {year} {2007})\BibitemShut {NoStop}%
\end{thebibliography}%

\end{document}